**TITLE: Deep learning enables high-throughput analysis of particle-aggregation-based bio-sensors imaged using holography**

**AUTHORS:** Yichen Wu[1,2,3,†], Aniruddha Ray[1,2,3,4,†], Qingshan Wei [1,2,3,#], Alborz Feizi[1,2,3], Xin Tong[1], Eva Chen[2], Yi Luo[1,2,3], Aydogan Ozcan[1,2,3,4,*]

**AFFILIATIONS:**

[1]Electrical and Computer Engineering Department, University of California, Los Angeles, California 90095, USA

[2]Bioengineering Department, University of California, Los Angeles, California 90095, USA

[3]California Nano Systems Institute (CNSI), University of California, Los Angeles, California 90095, USA

[4]David Geffen School of Medicine, University of California, Los Angeles, California 90095, USA

[†]Equal contribution authors

[#]Current address: Department of Chemical and Biomolecular Engineering, North Carolina State University, Raleigh, North Carolina 27695

[*]Corresponding author: ozcan@ucla.edu

**ABSTRACT:**

Aggregation-based assays, using micro- and nano-particles have been widely accepted as an efficient and cost-effective bio-sensing tool, particularly in microbiology, where particle clustering events are used as a metric to infer the presence of a specific target analyte and quantify its concentration. Here, we present a sensitive and automated readout method for aggregation-based assays using a wide-field lens-free on-chip microscope, with the ability to rapidly analyze and quantify microscopic particle aggregation events in 3D, using deep learning-based holographic image reconstruction. In this method, the computation time for hologram reconstruction and particle autofocusing steps remains constant, regardless of the number of particles/clusters within the 3D sample volume, which provides a major throughput advantage, brought by deep learning-based image reconstruction. As a proof of concept, we demonstrate rapid detection of herpes simplex virus (HSV) by monitoring the clustering of antibody-coated micro-particles, achieving a detection limit of ~5 viral copies/µL (i.e., ~25 copies/test).

Particle clustering-based assays have been regularly used for various biosensing applications ranging from the detection of small molecules[1] to proteins,[2,3] nucleic acids,[4,5] as well as metal ions.[6] In this type of assay, the particles, either micro- or nano- sized, are surface-functionalized with e.g., antibodies, aptamers or other macromolecules that are reactive towards a certain small molecule, protein, nucleic acid etc. In the presence of the target analyte, the particles generally aggregate to form clusters of different sizes, depending on the analyte concentration. This aggregation causes changes to the physical properties of the solution, which can be used to infer the presence of the target. Different type of particles synthesized from noble metals, carbon or polymers, both in micro- and nano- scale, can be used as the building blocks for these assays. Among others, latex polymer based assay in particular is frequently used in clinical settings to detect antigens or antibodies in bodily fluids.[7]

By counting the clustering events of the assay, the presence as well as the concentration of the target analyte in solution can be inferred. This read-out mechanism of the assay is performed according to the selection of the particle material and the target of interest. For example, noble metal particles have unique plasmonic properties that impart a distinct color to them, so their clustering-induced shape and size change alters their light absorption and scattering properties.[1,8–12] Thus the aggregation process for such plasmonic particles can be monitored through their color changes using spectrometers or other colorimetric methods.[1,9] Some of these aggregation assays can be monitored by bare eye at high concentration for their color changes. Equivalently, polymer-based colorless assays can be monitored through the change of optical opacity. However, these *macro-scale* visual inspection methods infer bulk-effects and therefore are generally not very sensitive as the results can be compromised by e.g., external environmental conditions such as temperature, ambient light, and/or inspector biases. Ideally, the individual clustering event in the assay should be monitored and counted using different types of microscopy, so that the size and shape of each cluster can be accurately characterized.[13–15] However, this process is usually laborious and slow. Moreover, microscopes used in those processes have a limited field-of-view (FOV) and depth-of-field (DOF), making it difficult to characterize a large sample volume at micro-scale.

Here, we demonstrate an automatic, high-throughput and real-time aggregation-based bio-sensor analysis method, covering a large number of micro-particles in 3D using deep learning assisted lens-free digital holographic microscopy.[16–19] As shown in Fig. 1, a partially coherent light source is used to illuminate the sample from above, and a hologram of all the clustered as well as individual particles is recorded in a single snapshot, over a large FOV of ~21 mm$^2$. Using a deep learning based holographic reconstruction and image processing workflow (Fig. 2), the clusters in this assay are reconstructed, auto-focused and counted rapidly in less than one minute. Quite importantly, the computation time required for hologram reconstruction and particle autofocusing steps (which forms the majority of the overall time budget), is independent of the number of particle clusters within the 3D sample volume, which is a major throughput advantage created by deep learning-based hologram reconstruction. As a proof of concept, based on this approach we developed a test for detecting the herpes simplex virus (HSV) using micro-particle clustering and automatically inferred the viral concentration with high accuracy. HSV is one of the most widespread viral infections that is estimated to have affected more than 50% of the adults in the US.[20,21] Using this method, we achieved a detection limit of ~ 5 cp/μL or 25 cp per test (i.e.,~ 1.16 cp per mm$^2$ imaging FOV), providing a clinically relevant level of sensitivity for HSV detection.[16] This lens-free-imaging-based particle clustering analysis and bio-sensor framework, enhanced by deep learning, provides a high-throughput, automatic, and rapid sensing performance. A similar approach can also be used for readout and quantification of other particle aggregation-based assays for the detection of proteins, nucleic acid and/or metal ions, and can easily be integrated with microfluidics and lab-on-chip platforms for high-throughput and rapid testing of samples in field settings.

**RESULTS AND DISCUSSION**

Fig.2 shows the image processing steps that we used to reconstruct and count micro-particle clusters. After capturing the whole FOV hologram, it was digitally back-propagated to 500 µm above the sensor surface, which was the rough distance from the image sensor plane to the sample on the coverslip (see Supplementary Fig. S1 for details). Using a convolutional neural network (CNN)[19], that was trained using pairs of randomly back-propagated holograms and in-focus phase recovered reconstructions, the micro-particle clusters were auto-focused and phase recovered in a single step, and all in parallel. Individual regions containing micro-particle clusters were extracted based on thresholding on the reconstructed image, which was followed by a virtual digital phase contrast (DPC) step. Locations of individual micro-particles in the cluster was marked using Laplacian of Gaussian (LoG) feature-based localization (see the Materials and Methods section for details).

Fig. 3 demonstrates the reconstruction of the micro-particle clusters of different sizes using this CNN-based reconstruction approach. As shown in the CNN reconstruction (third column of Fig. 3), the CNN corrects defocus aberration and simultaneously removes the twin image and background interference related artifacts in the back-propagated images (second column of Fig. 3), and successfully generates an in-focus, phase recovered reconstruction. This is especially important for large FOV imaging of a sample cartridge since different particle clusters within the sample FOV will have different depths, requiring autofocusing to individual particles. Thanks to deep learning based hologram reconstruction, this simultaneous auto-focusing and phase recovery process is automatically performed for ***all*** the micro-particle clusters in this FOV in ***one step***, i.e., with a fixed processing time. Stated differently, the processing time remains constant (~ 20 s) regardless of the number of particle clusters in the FOV, which is a major advantage enabled by deep learning based image reconstructions. Following this hologram reconstruction, the micro-particle morphology inside each cluster is better highlighted through the DPC process. Finally, using the LoG feature detector, individual micro-particles inside each cluster were correctly localized and counted, as marked by the red circle in the last column of Fig. 3. The whole process takes less than one minute (as detailed in Fig. 3), which provides rapid read-out of this micro-particle aggregation assay using deep learning.

As a proof of concept, we developed an aggregation assay for sensing HSV-1 particles. This virus binds to the antibodies attached to the micro-particles, causing the micro-particles to cluster. To characterize the system, we prepared solutions with different concentrations of HSV and mixed them with the cluster assay (see Materials and Methods section for details). Fig. 4 (a-f) shows the cluster size distribution for six different virus concentrations: 0, 6.25, 12.5, 25, 50 and 100 cp/µL. Three independent measurements were recorded for each concentration. As the virus concentration in the mixture solution increased, we observed a drastic increase in the percentage of the clustered micro-particles in the sample solution. As expected, we observed bigger and a larger number of clusters at higher virus concentrations. At very high concentrations of 50 cp/µL (Fig. 4(e)), and 100 cp/µL (Fig. 4(f)), we note a significant increase in larger clusters (with ≥3 particles in each), whereas the percentage of smaller clusters (with two particles) increased only slightly (see Supplementary Fig. S2 for details). For reporting the sensor output, counting the ratio of each cluster size is more reliable than counting their absolute numbers, as it mitigates the errors due to nonuniform distribution of the micro-particles inside the solution, pipetted onto the test/sample coverslip.

Quantification of micro-particle clustering can also be used to determine the virus concentrations by analyzing the number of clusters and their respective size, creating a calibration curve. In order to demonstrate this opportunity, we used a new quantity "clustered particles", defined as the number of particles inside each cluster minus one (to cancel out the signal for individual unbound particles). If we consider the individual particles as the nodes in a graph, this quantity is equivalent to the minimal number of edges that could link these particles together (i.e., a tree structure). Based on this analysis, Fig. 5 shows the ratio of the clustered particles to the total number of particles inside the imaging FOV, for different virus concentrations. This ratio exhibits a linear relationship as a function of the viral concentration for relatively lower concentrations; a saturation effect towards the higher concentrations is also observed, which defines the upper limit of the viral load quantification range for this assay. This saturation at high viral concentrations is due to the limited amount of active microparticles in the assay that can form clusters in the solution.

To quantify the detection limit of the system for HSV sensing, we took the data of the first four concentrations and fit a linear regression model, as shown in Fig. 5(b). Following this linear fit, the detection limit was calculated as the mean plus three times the standard deviation of clustered particles in the control sample (0 cp/μL), which was estimated to be ~5 cp/μL, i.e., 25 viruses per test (where a 5 μL droplet is imaged per test), as marked by the dashed line in Fig. 5(b). This detection limit is clinically relevant as HSV (HSV-1 and HSV-2) concentration in patient samples varies between 0-$10^5$ virus cp/μL, depending on the method of collection and the phase of the outbreak.[22] Therefore, this method has the potential for early detection of HSV infection, especially when the viral copy number is generally low and the traditional clinical tests rely on viral culture methods to enhance the concentration. [23]

Here we used a micro-particle-based assay due to its inherent larger particle size and high particle scattering cross section that can be easily and rapidly detected using a single snapshot hologram captured with a simple and cost-effective lens-free on-chip microscope. However, this method can also be easily adapted for imaging and characterizing sub-micron sized particles by implementing techniques developed for lens-free on-chip microscopes such as holographic pixel super-resolution,[24–26] polymer nano-lenses,[27–29] and/or using ultraviolet (UV) illumination.[30] Another advantage of our presented method is the ability to rapidly characterize a wide range of micro-particle concentrations, even in 3D, including low concentrations, unlike the bulk-effect visual techniques which require extremely high concentrations of micro-particles. Utilizing a smaller number of micro-particles can reduce the cost of an assay significantly due to the smaller number of antibodies and/or other expensive chemicals required for surface functionalization.

**CONCLUSION:**

We developed a fast, sensitive and automated read-out method for particle clustering-based assays using wide-field lens-less computational on-chip microscopy powered by deep learning. A CNN-based algorithm is used to reconstruct the micro-particle clusters in 3D, all in parallel, and facilitate the detection and counting of the particles inside each cluster using a single hologram snapshot. Based on DPC and LoG feature detector, individual micro-particles inside each cluster are accurately localized and analyzed in less than one minute. As a proof of concept, we demonstrated accurate and rapid read-out of an HSV micro-particle clustering assay with a detection limit of ~5 cp/µL. This method can also be widely applied to rapidly characterize other types of particle clustering-based bio-sensor assays.

## METHODS

**Materials**

The streptavidin coated silica particles (~2 µm diameter) were purchased from Microspheres-Nanospheres Inc. (item # C-SIO-2.0SAv). The biotin tagged Herpes Simplex Virus type 1/2 polyclonal antibody and Phosphate Buffered Solutions (PBS) was acquired from Thermo Fischer Scientific (item # 10010023). The HSV viral culture (NATtrol Herpes Simplex Virus Type 1 Strain: MacIntyre, 50,000 cps) was acquired from ZeptoMetrix Corporation. All the chemicals and materials were used as received.

**Sample Preparation**

The particles (20 µL) used for the experiments were first washed by diluting in PBS (200 µL) followed by centrifugation. The supernatant was discarded and replaced with pure PBS buffer. This step was performed three times and the particles were finally suspended in 400 µL of PBS. This washing step is important as the viral buffer itself can also cause clustering of the micro-particle assay, leading to false detections of virus. The biotin coated antibodies were added to the particles (ratio of 1:55 v/v) and incubated for 1.5 hours. Following incubation of the unattached antibodies were washed away by centrifugation. The antibody coated particles were re-suspended in PBS at a concentration of $1.5 \times 10^4$ particles/mL. This antibody coated particles can be used immediately for experiments or stored in the freezer for up to a week. The virus solution at different concentration was added to the particles at equal volumes and mixed overnight (~16 hours) and then used for imaging. Prior to adding the particles, the viruses were isolated from their original solution using a centrifuge filter and re-suspended in fresh PBS.

    The coverslips were thoroughly cleaned before imaging. They were first incubated in acetone overnight and then washed with ethanol and water. The coverslips were plasma treated for 120 seconds prior to using them.

**Lens-free microscopy**

The lens-free microscope utilized a partially coherent source ($\lambda = 510$ nm, $\Delta\lambda = 3$ nm, coupled to a single mode fiber) as illumination, and a CMOS image sensor chip (Sony IMX081, 1.12 µm pixel size) to capture holograms of the samples (see Fig. 1 for details). A clean, plasma-treated #1 coverslip (thickness of ~150 μm) was placed directly on top of the image sensor chip for analysis. A droplet (5μL) of mixture solution was pipetted directly onto the center of this coverslip. A hologram image was recorded immediately thereafter. Three independent measurements were taken for each sample.

**Hologram image pre-processing and reconstruction**

The image sensor used was a Bayer color image sensor. To correct the non-uniform illumination and sensor response on the captured hologram, we performed a wavelet transform (using order-eight Symlet [31]) on the two green Bayer channels of the sensor individually, extracted sixth-level approximation as the background shade, and divided each green channel output value by this background shade. Then, the pixels in these two green channels were re-arranged back to their original locations, and the red and blue pixels were linearly interpolated from their adjacent green pixels. This step generates a *shade-corrected hologram*, where the four Bayer channels and illumination shade are equalized, and the resulting image has a uniform background centered at one.

This corrected hologram was then digitally free-space back-propagated using:

$$\text{ASP}[I(x,y); \lambda, n, -z_2] = 1 + o(x,y) + t(x,y) + s(x,y) \tag{1}$$

where $\lambda = 510$ nm is the illumination wavelength, n = 1.5 is the refractive index between the sample and the image sensor, $z_2$ = 500 μm is the rough spacing between the sample to the sensor plane, which was initially based on the physical thickness of the glass coverslip and the protective filter of the image sensor chip, and also validated to be the center of the depth distribution of micro-particles in the 3D volume (see supplementary Fig. S1 for details). ASP[.] is the angular spectrum propagator, which is calculated by taking the Fourier transform of the hologram, and multiply it by the angular spectrum filter, followed by an inverse Fourier transform back to the spatial domain. The hologram was padded 2× in the Fourier domain. The exact value of $z_2$ is **not** critical for our approach as there is an additional auto-focusing step, which will be detailed next. To speed up the image processing step, this $z_2$ distance was chosen globally without finding an individual focus for each cluster. This defocus aberration for individual particle clusters (e.g. second column of Fig. 3) was corrected rapidly (and all in parallel for all the particles/clusters) by using a CNN-based simultaneous auto-focusing and phase recovery process, discussed in the next subsection.

**CNN based micro-particle cluster reconstruction**

A CNN was trained to perform simultaneous autofocusing and phase recovery for each micro-particle cluster within the back-propagated hologram FOV[19]. To train this CNN, a series of measurements were made, where the micro-particle solution was confined between two coverslips to prevent evaporation. After waiting for 10 min for all the micro-particles to sediment onto the bottom layer of the coverslip, a hologram image was taken using the same setup. This image is cropped into non-overlapping patches of 512-by-512 pixels. Auto-focusing[32–34] and an object support iterative phase recovery algorithm[35–37] were used on these individual hologram patches to generate a target image for the network, and these original hologram patches were randomly back-propagated to +/- 200 μm around the focal plane, which were used as the input image to the network. This defocus range to train the CNN covers all the particle clusters in the FOV (see e.g. Supplementary Fig. S1). The CNN has the same architecture as described in Ref. 19, and is built on TensorFlow.[38] Similar CNNs have been used for image segmentation,[39] image super-resolution,[40,41] and image quality enhancement.[42–44] After training for ~ 100 epochs, the CNN learns to auto-focus and phase-recover particle clusters at the same time. Note that in this training data set, only images of micro-particles that locate on the same plane were used. However, thanks to the convolutional nature of a CNN, the network learned to simultaneously auto-focus and phase-recover each individual micro-particle cluster all in parallel, saving the need to focus and reconstruct each individual micro-particle cluster separately, which significantly boosts the computational efficiency for the detection and counting of the particle clusters. We emphasize that this training process is only done once, and after its training the CNN remains fixed and is blindly applied on new holograms.

**Detection of micro-particles using DPC and LoG feature detector**

To enhance the signatures of individual micro-particles inside a cluster before the detection step, the DPC processing step was applied, where we took the Fourier transform of the reconstructed complex region-of-interest and shifted the zeros frequency component by -π/2, followed by an inverse Fourier transform.

A LoG filter with window size of 7-by-7 pixels and $\sigma = 0.6\mu m$ (which effectively targets circular blobs with a diameter of d = $2\sqrt{2}\,\sigma \approx 2$ μm) was applied on the DPC image. The local maxima with a pixel value above the threshold (0.1) were detected. Non-maximum suppression was used to eliminate the local maxima that were too close to each other (i.e., with a distance <1.5 μm), to avoid multiple detection of the same micro-particle. This number was used to estimate the size for each cluster and generated the corresponding statistics in Fig. 4 and Fig. 5.

**ASSOCIATED CONTENT**

**Supporting Information**

Microbead cluster height distribution on the lens-less image. Assay composition change due to virus concentration change.

**AUTHOR INFORMATION**

**Author Contributions**

A.O, Y.W., A.R. and Q.W. initiated the research. A.R., A.F. and E.C. prepared the virus assay. Y.W., A.R., A.F. and E.C. did the imaging experiments. Y.W. wrote the data processing program. X.T. combined and refined the processing program. Y.W. and X.T. analyzed the experiment data. Y.L. contributed to the experiments. A.O., Y.W. and A.R. prepared the manuscript. A.O. supervised the research.

**ACKNOWLEDGEMENTS**

The Ozcan Group at UCLA acknowledges the support of the National Science Foundation, NSF Partnerships for Innovation: Building Innovation Capacity (PFI: BIC) Program, Vodafone Americas Foundation, and the Howard Hughes Medical Institute. We acknowledge Dr. Hyouarm Joung, Ms. Farharna Haque, Ms. Courtney Chang, and Mr. Yicheng Li of UCLA for their help during the optimization process of the sample preparation. AR acknowledges the Ruth L. Kirschstein Institutional National Research Service Award (5T32DK104687-03) from NIH.

**Figures and captions**

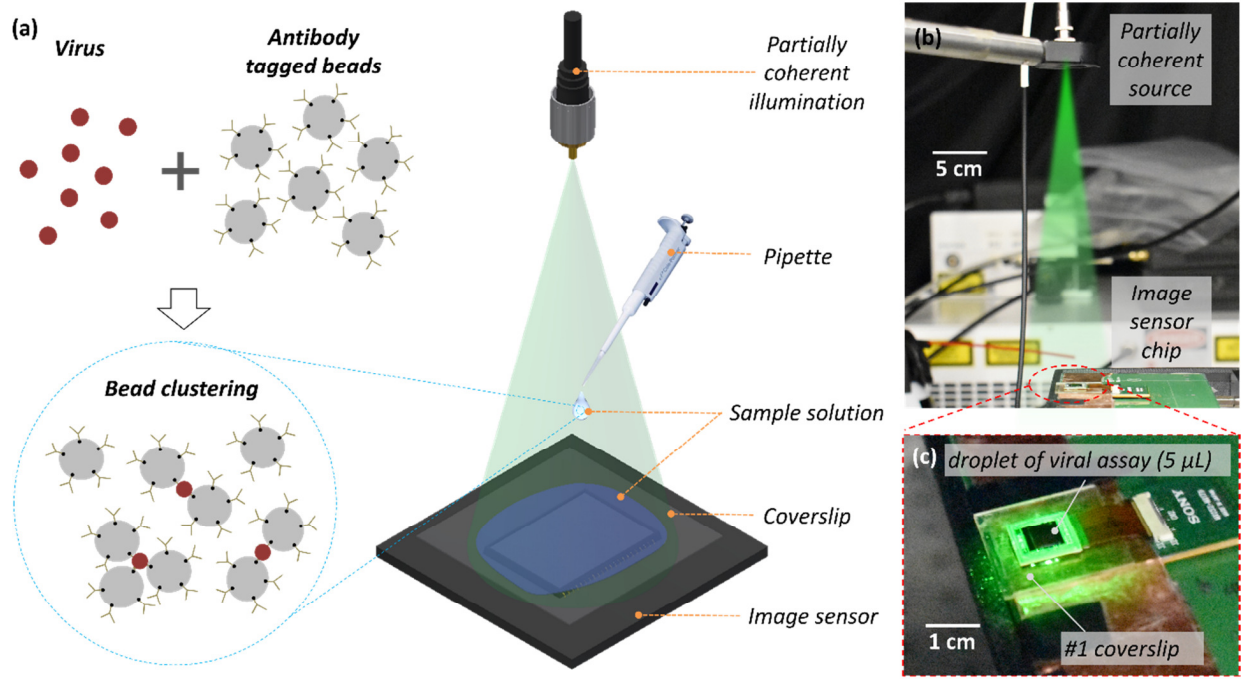

**Fig. 1. Deep learning-based holographic imaging enables high-throughput analysis of particle-aggregation-based bio-sensors.** (a) Schematics of the assay. The solution containing the virus particles of interest was mixed with antibody tagged micro-particles, forming micro-particle clusters in the sample solution. The sample solution was then pipetted onto a plasma-treated coverslip. An in-line hologram was captured immediately after the pipetting step. (b) Photo of the lens-less imaging setup. (c) Zoomed-in region of (b).

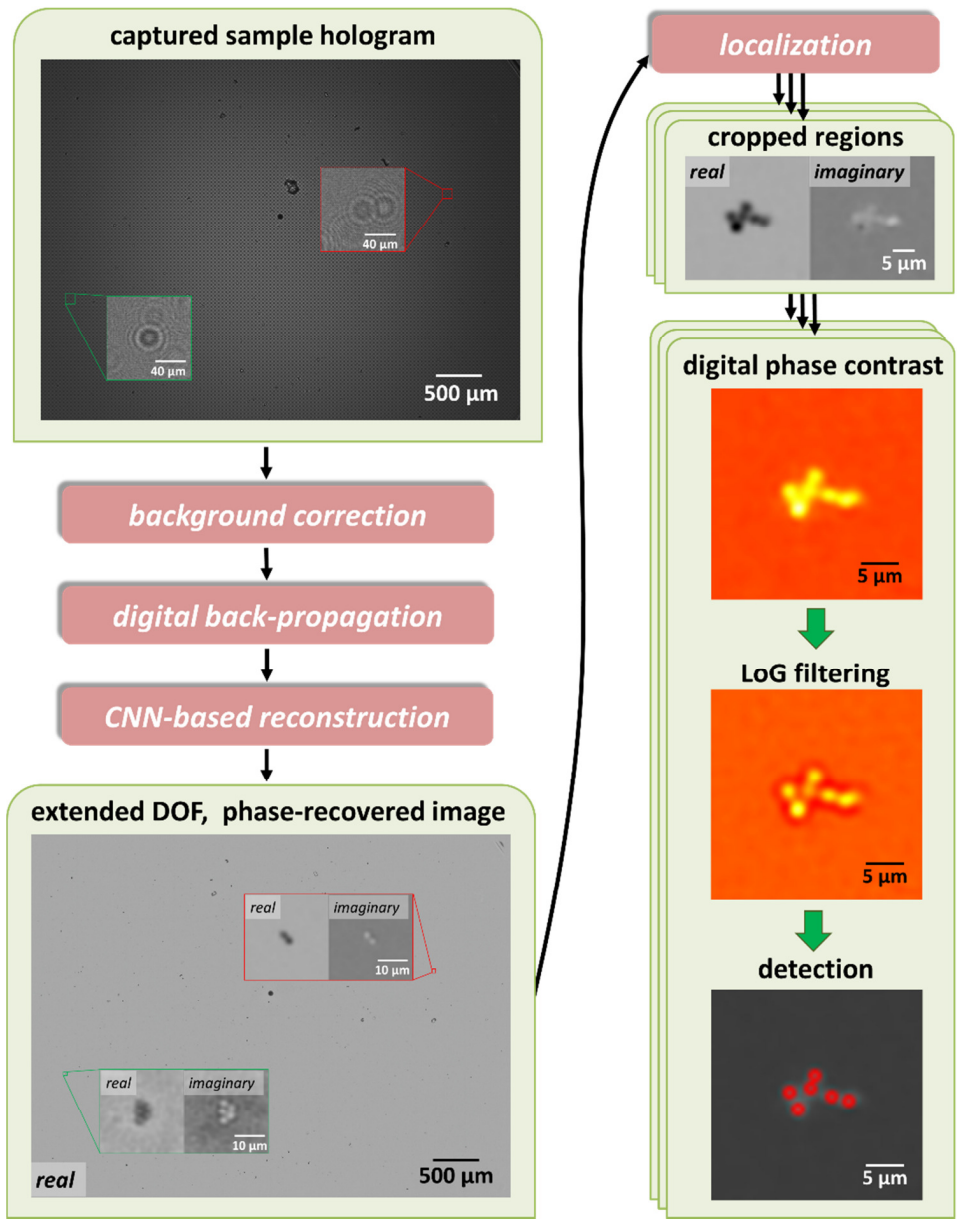

**Fig. 2. Image processing of micro-particle clusters.** A lensfree hologram of the micro-particle cluster sample was captured. Using a CNN, micro-particle clusters were simultaneously auto-focused and phase recovered in a single step and all in parallel. Individual micro-particle cluster images were cropped and analyzed rapidly from the reconstructed image.

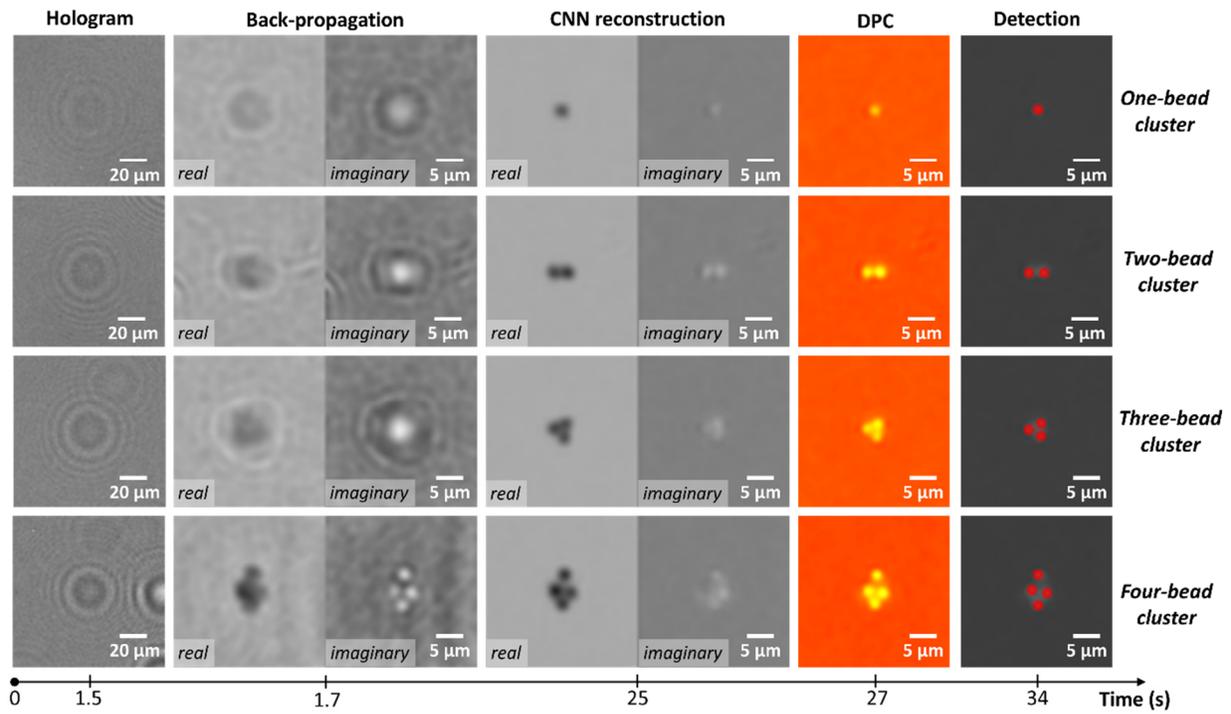

**Fig. 3. CNN-based reconstruction of micro-particle clusters.** Using a trained CNN, each individual particle cluster within the sample volume was focused onto the same plane and phase recovered in parallel. This enables rapid detection of each micro-particle cluster in the image without the need to search their correct focus in 3D. Cropped regions with one, two, three and four micro-particles are shown, as examples. The time axis at the bottom marks the processing time for *full* FOV (containing ~ 1,200 particle clusters over a FOV of ~21 mm$^2$) after each one of the image processing steps. t=0 marks the acquisition of the hologram.

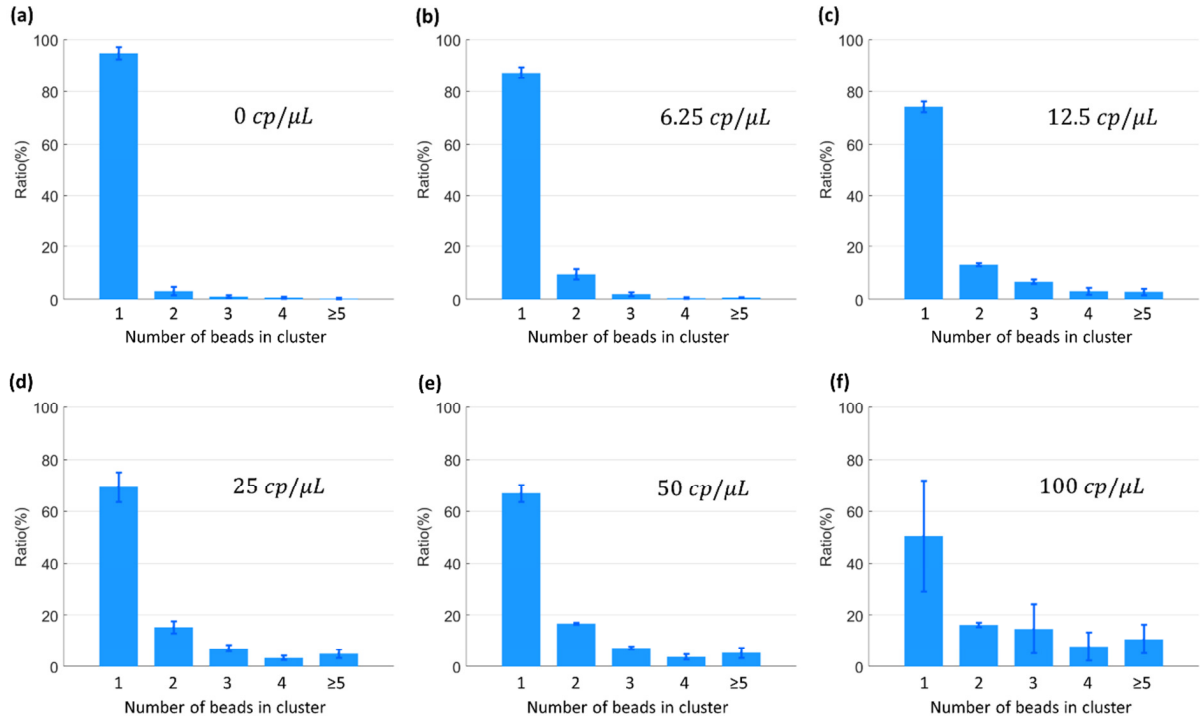

**Fig. 4. Histograms of the detected cluster size as a function of the viral concentration.** Concentrations in (a-f) were 0, 6.25, 12.5, 25, 50, and 100 cp/μL, respectively. Three measurements were taken for each measurement.

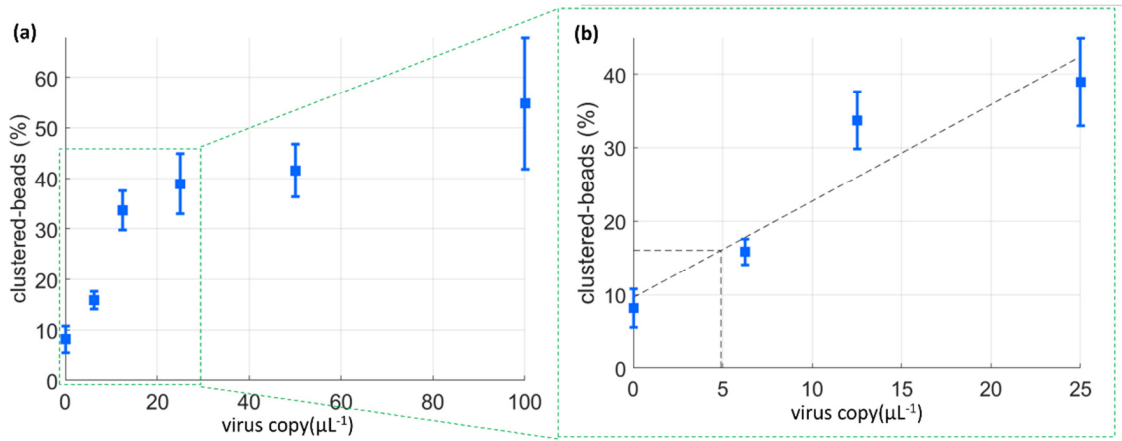

**Fig. 5. Titration curve for quantifying virus concentration using our particle clustering-based assay.** The percentage of the clustered micro-particles with respect to the total number of micro-particles in the solution was used to infer the viral concentration. Linear regression was used for the first four data points, resulting in Y [%] =1.31 X [cp · µL$^{-1}$] + 9.61. Following this fit, the detection limit was calculated to be ~4.91 viral cp/µL, i.e., ~25 viruses per test, as marked by the dashed line. Three experiments were performed for each viral concentration.